\documentclass[aps,prb,showpacs,twocolumn]{revtex4-1}

\usepackage{amsmath}
\usepackage{amssymb}
\usepackage{graphicx}

\begin{document}

\title{Diffusion coefficients from signal fluctuations:\\ Influence of
  molecular shape and rotational diffusion}

\author{Susanne Hahne} \affiliation{Fachbereich Physik, Universit\"at
  Osnabr\"uck, Barbarastra{\ss}e 7, 49076 Osnabr\"uck, Germany}

\author{Philipp Maass}
\email{philipp.maass@uni-osnabrueck.de}
\affiliation{Fachbereich Physik, Universit\"at Osnabr\"uck,
  Barbarastra{\ss}e 7, 49076 Osnabr\"uck, Germany}

\date{December 9, 2013}

\begin{abstract}

  Analysis of signal fluctuations of a locally fixed probe, caused by
  molecules diffusing under the probe, can be used to determine
  diffusion coefficients. Theoretical treatments so far have been
  limited to point-like particles or to molecules with circle-like
  shapes. Here we extend these treatments to molecules with
  rectangle-like shapes, for which also rotational diffusion needs to
  be taken into account.  Focusing on the distribution of peak widths
  in the signal, we show how translational as well as rotational
  diffusion coefficients can be determined. We address also the
  question, how the distribution of interpeak time intervals and
  autocorrelation function can be employed for determining diffusion
  coefficients. Our approach is validated against kinetic Monte
  Carlo simulations.

\end{abstract}






\maketitle

\section{Introduction}\label{sec:intro}

Self-assembly of molecules on surfaces has become a topic of intensive
research, both from a fundamental point of view and for developing
electronics based on molecular units.\cite{Rahe/etal:2013,
  Hlawacek/Teichert:2013, Einax/etal:2013, Kuehnle:2009,
  Kowarik/etal:2008} For controlling the kinetic growth, the diffusion
coefficient is an important parameter. Various techniques are used to
measure molecular diffusion on surfaces.\cite{Barth:2000} For fast
diffusion, a particularly suited method is an analysis of signal
fluctuations caused by molecules entering and leaving the detection
area under a probe. Corresponding time series can be recorded by a
scanning tunneling microscope, \cite{Lozano/Tringides:1995} or, in
principle, also by an atomic force microscope or some other probe with
a point like sensor. By fixing the probe locally above the substrate,
the time resolution can be significantly increased, thus allowing to
capture high mobilities, as often encountered in the case of molecular
motion.

Different means are available for evaluating the detection events in
the time series. Originally, the autocorrelation function (ACF) has
been the focus of experimental \cite{Tringides/etal:1999} and
theoretical \cite{Sumetskii/Kornyshev:1993} studies. For atomic
diffusion, the ACF method yields relative changes of diffusion
coefficients $D$. Recently, this method has been extended to molecules
with sizes larger than the step length of translational
moves.\cite{Hahne/etal:2013} This way absolute values of $D$ become
accessible. Furthermore, the residence
time-distribution\cite{Ikonomov/etal:2010} (RTD) has been discussed
and another new method, the interpeak-time distribution (ITD), has been
introduced. The RTD is the distribution of peak widths and the ITD is
the distribution of time intervals between successive peaks. Both
of them are suitable to determine absolute values of $D$. 

Various
merits and limits have been discussed for the three variants based on
the ACF, RTD, and ITD.\cite{Hahne/etal:2013} From a general point of
view, inter-molecular and molecule-tip interactions can affect the
results. The ACF is influenced by both of these interactions, while
the ITD is influenced only by inter-molecular
interactions. Accordingly, the RTD method can be viewed as the most
powerful one, because it is only influenced by the molecule-tip
interaction, which can be reduced systematically by adjusting the
molecule-probe distance.\cite{Ikonomov/etal:2010}

\begin{figure}[b]
 \centering
 \includegraphics[width=8cm]{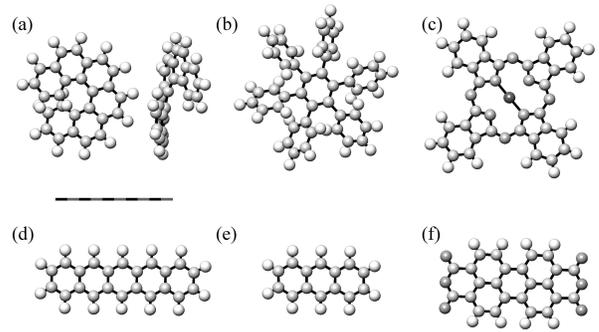}
 \caption{Ground-state vacuum structures obtained from density
   functional calculations of several molecules. These molecules can
   be viewed as representative for classes of derivatives that are
   widely used in studies of molecular self-assembly on surfaces: (a)
   helicene\cite{Seibel/etal:2013} (top and side view), (b)
   hexaphenylbenzene,\cite{Gross/etal:2005} (c) (copper-)
   phtalocyanine,\cite{Mugarza/etal:2012} (d)
   pentacence\cite{Gross/etal:2009} (e)
   anthracene,\cite{Wan/Itaya:1997, Potapenko/etal:2010} and (f) the
   perylene derivate PTCDA.\cite{Paulheim/etal:2013} Images of the
   molecules in the respective studies suggest that their overall
   shape is not significantly distorted upon adsorption and that they
   often lie flat on the surface. Circular and rectangular shapes can
   hence be expected to describe their detection area under the probe
   to a good approximation.}
 \label{fig:footprints}
\end{figure}

So far, the focus lay on molecules, whose shape can be effectively
described by a circle. Examples of such circular-type molecules,
widely used in present studies of self-assembly on surfaces, are shown
in Fig.~\ref{fig:footprints}(a)-(c). Other molecules in such studies,
as shown in Fig.~\ref{fig:footprints}(d)-(f), are better described by
rectangles.  In this work, we extend our theoretical treatment to
these shapes.  Rectangular-type molecule render signal fluctuations
from rotational diffusion possible. Including such rotational moves in
the theoretical description increases the complexity significantly. We
offer a first approach to this problem, considering thermally
activated rotational diffusion uncoupled from translation. Our focus
in this work will be on the RTD, because of its advantage over the
other methods. In the last section~\ref{sec:conclusions}, we will also
adress the question how rectangular shapes and rotational diffusion
affect the ACF and ITD. As a further point, we discuss in
Sec.~\ref{sec:isotropic} the question how many peaks need to be
recorded to obtain faithful results for the diffusion coefficient.

\section{KMC simulations}\label{sec:kmc}

The quality of approximations entering our analytical treatments is
checked by comparison with results obtained from kinetic Monte Carlo
(KMC) simulations. These simulations are carried out on a square
lattice with lattice constant $a=1$ (setting our length unit) and
periodic boundary conditions. The center positions of $N$ particles of
circular or rectangular shape perform jumps between nearest neighbor
lattice sites with rate $w_D=4D/a^2=1$ ($w_D^{-1}$ setting our time
unit), where $D$ is the diffusion coefficient. The latter is to be
compared to $\tilde D$ values extracted from the analysis of simulated
data, which serve as surrogate for experimental ones. The particles
represent freely diffusing molecules in an experiment and are assumed
to have a coverage much smaller than a monolayer. We are neglecting any
interaction effects between them. When including rotational dynamics,
reorientation moves with rate $w_\varphi$ are additionally taken into
account, see Sec.~\ref{sec:rotation}. Translational and rotational
dynamics are implemented by using the reaction time
algorithm.\cite{Holubec/etal:2011}

\begin{figure}[b]
 \centering
 \includegraphics[width=8cm]{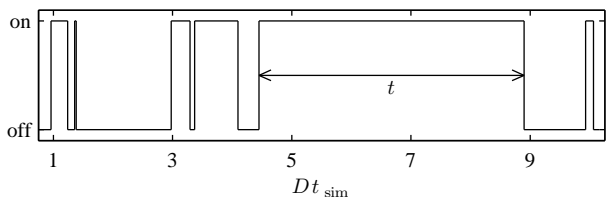}
 \caption{Part of a simulated ``on-off'' time series. Peak widths
   are denoted by $t$.}
 \label{fig:signal}
\end{figure}

Time series are generated by selecting one fixed lattice site, the
so-called ``probe-site'', which represents the point on the surface,
from which the signal in an experiment is recorded. The simulated
signal is ``on'' if the probe-site is covered by a part of a molecule,
while otherwise it is ``off''.  The thus-generated ``on-off'' time
series resemble measured ones, if in the latter noise is eliminated
and a proper threshold is set, which allows for a reduction to a
rectangular signal. The details of this procedure are described in
Ref.~\onlinecite{Hahne/etal:2013}. The example shown in Fig.~2(b) of
that work may be compared to the simulated signal displayed in
Fig.~\ref{fig:signal}.

\section{RTD for molecules with circular shapes}
\label{sec:isotropic}

In the following we shortly summarize our previous results from
Ref.~\onlinecite{Hahne/etal:2013} with respect to the RTD, because we
will refer repeatedly to the corresponding equations in the following
sections. The methods were developed for molecules, with size
large compared to the sensitive area of the probe. This way the probe
can be considered as pointlike and the molecule extension gives rise
to a detection area, which is defined by the set of molecule center
positions around the tip, that will cause a detectable change in the signal.  For
molecules with approximately circular shapes, as the ones shown in
Fig.~\ref{fig:footprints}(a)-(c), a circle with radius $R$ can be
assigned to the detection area, for example, by taking its gyration
radius. The RTD then is given by
\begin{equation}\label{eq:psiRTD}
  \Psi_{\rm circ}(t)=
\frac{2D}{R^2}\sum_{n=1}^\infty 
\frac{x_n~J_0(x_n (1-\frac{\Delta}{R}))}{J_1(x_n)}\,
\textrm{e}^{-x_n^2 Dt/R^2}\,,
\end{equation}
where $J_{\nu}(.)$ is the Bessel function of first kind with $x_n$ its
$n$-th zero; $\Delta$ is the minimal penetration depth of the center
into the detection area under the tip, which is at least necessary to
turn the signal ``on''. For times much larger than the typical time
$\tau_R\equiv R^2/(x_1^2 D)$ for the molecule to explore the detection
area,
\begin{equation}\label{eq:psiRTDII} 
  \Psi_{\rm circ}\left(t\right)\sim
\frac{2D}{R^2}\frac{x_1~J_0(x_1(1-\frac{\Delta}{R}))}{J_1(x_1)}\,
\textrm{e}^{-x_1^2Dt/R^2}\,,  
\end{equation}
while a power law $\sim t^{-3/2}$ characterizes the behavior
for $\Delta^2/D\ll t \ll \tau_R$. For
$t\lesssim\Delta^2/D$ the continuum description breaks down.

\begin{figure}[b]
 \includegraphics[width=8cm]{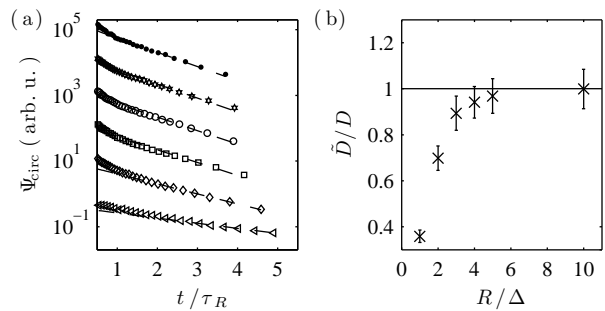}
 \caption{(a) Simulated RTDs for various fractions $R/\Delta=1$
   ($\lhd$), 2 ($\diamond$), 3 ($\square$), 4 ($\circ$), 5 ($\star$),
   and 10 ($\bullet$). The dashed lines are least square fits to
   Eq.\,(\ref{eq:psiRTDII}). The distributions were vertically shifted
   for better visibility. (b) Diffusion coefficients $\tilde D$ from
   the fitting, divided by the input value $D$ in the KMC simulation,
   as a function of $R/\Delta$.}
\label{fig:extended}
\end{figure}

Application of Eq.~(\ref{eq:psiRTD}) and its derivates requires
$R/\Delta$ to exceed a certain threshold, because for $\Delta$ close
to or larger than $R$, one would need to consider molecular jumps into
and out of the detection area instead of a diffusive motion within
this area. The jump motion could be treated as well, but this is not
our focus here. Corresponding events may be difficult to detect within
experimental time resolution. Nevertheless, before starting to address
our main topics, it is interesting to see, down to which fractions
$R/\Delta$ the continuum treatment gives reliable results for
$D$. Simulated RTDs for various fractions $R/\Delta$ are shown in
Fig.~\ref{fig:extended} and were fitted with
Eq.~(\ref{eq:psiRTDII}). In these simulations we fixed $\Delta$ and
$D$, and varied $R$. The fits [dashed lines in
Fig.~\ref{fig:extended}(a)] yield estimates $\tilde D$ in good
agreement with $D$ for $R/\Delta\gtrsim3$, see
Fig.~\ref{fig:extended}(b).

\begin{figure}[t]
 \centering
 \includegraphics[width=8cm]{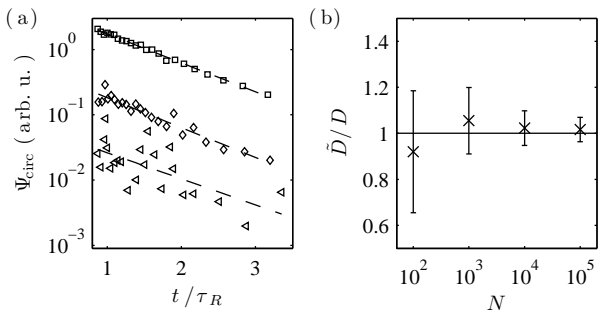}
 \caption{(a) Simulated RTDs for various total numbers of events $N=$
   10$^3$ ($\lhd$), 10$^4$ ($\diamond$), and 10$^5$ ($\square$). The
   dashed lines are least square fits to Eq.\,(\ref{eq:psiRTDII}), and the
   distributions were vertically shifted for better visibility. (b)
   $\tilde D/D$ as a function of $N$.}
\label{fig:NoEvents}
\end{figure}

Another essential factor for applications is how many events need to
be recorded for obtaining reliable results. The long-time regime
$t\gg\tau_R$ of the RTD is most relevant for the fitting, see
Eq.~(\ref{eq:psiRTDII}). From Eq.~(\ref{eq:psiRTD}) it can be
calculated that about 4\% of RTD events are in this
regime. Figure~\ref{fig:NoEvents}(a) shows RTDs for different total
numbers of events. Fits to the exponential decay according to
Eq.~(\ref{eq:psiRTDII}) are marked by the dashed lines. Already for in
total 10$^3$ events, i.e.\ about 40-100 events in the long-time
regime, $\tilde D$ is close to $D$, but, as shown in
Fig.~\ref{fig:NoEvents}(b), the error is quite large. Taking larger
number of events reduces the error. In view of the additional
uncertainties in experiments, we suggest to perform measurements
capturing at least 10$^4$ events in total.

\section{RTD for molecules with rectangular shapes}
\label{sec:molShape}
So far, a circular detection area with radius $R$ was assigned to the
molecules. We now extend this ansatz to molecules to which rectangles
can be assigned with a longer and shorter edge $L_{\rm l}$ and $L_{\rm
  s}$, corresponding to an aspect ratio $\alpha = L_{\rm l}/L_{\rm
  s}\geq1$. Only for $\alpha$ significantly larger than one this will
give rise to notable changes of the RTD compared to a description
based on a circular detection area.

An analytical expression for the RTD in case of rectangular shapes is
readily obtained. Considering a molecule center entering the
rectangular detection area and its diffusion within the area, we are
led to the problem of determining the diffusion propagator
$p\left(\mathbf{r},t\right)$ in the presence of a closed absorbing
boundary formed by the edges of the detection area.  A uniform
distribution of the molecule center on an inner rectangular contour,
displaced by $\Delta$ from the absorbing boundary, is used as initial
condition.

Expansion of $p\left(\mathbf{r},t\right)$ in terms of the
eigenfunctions of the Laplacian, $\chi_{m,n}= \sin[k^{\rm
  (l)}_mx]\sin[k^{\rm (s)}_ny]$, $k^{\rm (l)}_m=(2m+1)\pi/L_{\rm l}$,
$k^{\rm (s)}_n=(2n+1)\pi/L_{\rm s}$, $m,n=0,1,\ldots$, gives
\begin{equation}\label{eq:pRTDrec}
  p\left(\mathbf{r},t\right) =
\frac{8}{L_{\rm l}L_{\rm s}}\sum_{m,n=0}^\infty C_{m,n}\chi_{m,n}
  \exp(-k_{m,n}^2Dt)
\end{equation}
with $k_{m,n}^2=({k^{\rm (l)}_m}^2+{k^{\rm (s)}_n}^2)$,
and coefficients
\begin{equation}
C_{m,n} = 
\frac{\sin(k^{\rm (l)}_m\Delta)\cos(k^{\rm (s)}_n\Delta)}
{k^{\rm (s)}_n(L_{\rm l}+L_{\rm s}-4\Delta)}+
\frac{\sin(k^{\rm (s)}_n\Delta)\cos(k^{\rm (l)}_m\Delta)}
{k^{\rm (l)}_m(L_{\rm l}+L_{\rm s}-4\Delta)}\,.
\end{equation}
The RTD follows from
$\Psi_{\rm rec}(t) = -\partial_t
\int_0^{L_{\rm l}}\int_0^{L_{\rm s}} 
\mathrm{d}x\,\mathrm{d}y\,p\left(\mathbf{r},t\right)$, yielding
\begin{equation}\label{eq:psiRTDrec}
  \Psi_{\rm rec}(t) = \frac{32D}{\pi^2} 
  \sum_{m,n}\frac{C_{m,n}\,k_{m,n}^2\,
\exp\left(-D k_{m,n}^2t\right)}{(2m+1)(2n+1)}\,.
\end{equation}
Not surprisingly, this RTD shows the same functional time dependence
as the RTD for circular-shaped molecules discussed in
Sec.~\ref{sec:isotropic}, that means a power law decay at intermediate
times $\Delta^2/D\ll t\ll \tau_L\equiv1/k^2_{0,0}D$ and an exponential decay
\begin{equation}\label{eq:psiRTDrecII}
  \Psi_{\rm rec}(t)
\sim \frac{32D}{\pi^2}\, C_{0,0}\,k_{0,0}^2
  \exp\left(-D k_{0,0}^2t\right)
\end{equation}
for $t\to\infty$ ($t\gg \tau_L)$. Only the prefactors in
these laws change. 

\begin{figure}[t]
 \centering
 \includegraphics[width=8cm]{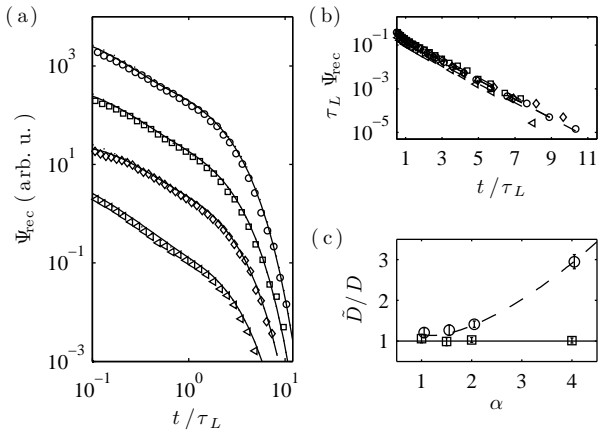}
 \caption{(a) Four simulated RTDs for molecules with rectangular
   shapes: $L_{\rm s}=5$, $L_{\rm l}=\alpha L_{\rm s}$, with $\alpha=1.0$ ($\lhd$), 1.5
   ($\diamond$), 2.0 ($\square$), and 4.0 ($\circ$). Curves were
   vertically shifted for better visibility.  The solid lines are best
   fits of Eq.~(\ref{eq:psiRTDrec}) to the data after determination of
   $\tilde D$. (b) Exponential decay of $\Psi_{\rm rec}$ in the
   long-time regime and least square fits (dashed lines) for
   determining $\tilde D$. (c) $\tilde D/D$ as a function of the
   aspect ratio $\alpha=L_{\rm l}/L_{\rm s}$.  Squares refer to
   fitting with Eq.~(\ref{eq:psiRTDrecII}), and circles to fitting
   with Eq.~(\ref{eq:psiRTDII}), with $R$ taken the gyration radius.}
\label{fig:recDistribution}
\end{figure}

RTDs obtained from KMC simulations of molecules with rectangular
shapes are depicted in Fig.~\ref{fig:recDistribution}(a). Fitting
Eq.~(\ref{eq:psiRTDrecII}) to the long-time behavior allows one to
determine estimates $\tilde D$ for the diffusion coefficients, as
demonstrated in Fig.~\ref{fig:recDistribution}(b).
Figure~\ref{fig:recDistribution}(c) shows that for aspect ratios close
to one, it does not make a significant difference whether the
simulated data are fitted with Eq.~(\ref{eq:psiRTDII}) (taking $R$ as
the gyration radius of the rectangle) or with
Eq.~(\ref{eq:psiRTDrecII}). For aspect ratios $\alpha\gtrsim1.5$,
however, a description in terms of circular shapes yields erroneous
diffusion coefficients.

After having determined $\tilde D$ from the long-time behavior,
Eq.~(\ref{eq:psiRTDrec}) can be used to fit $\Psi_{\rm rec}(t)$ for
all times (except very short ones, where the continuum treatment
becomes invalid).  Corresponding fits are shown by the solid lines in
Fig.~\ref{fig:recDistribution}(a). As a result, the parameter $\Delta$
can also be estimated, which is useful for a consistency check. It
should have a size of about the jump length, i.e.\ it generally should
be comparable to the lattice constant of the substrate. Evaluation of
the data in Fig.~\ref{fig:recDistribution}(a) yields
$\Delta\simeq0.7$, which is close to the jump length $a=1$ in our
simulations.

To a good approximation, the overall behavior can also be accounted
for by using the solution Eq.~(\ref{eq:psiRTD}) for circular-shaped
molecules with an effective radius $R_{\rm mod}$. This is obtained by
comparing the long-time limits in Eqs.~(\ref{eq:psiRTDII}) and
(\ref{eq:psiRTDrecII}), yielding
\begin{equation}
\label{eq:rmod}
R_{\rm{mod}} =\frac{x_1}{\pi}\left(L_{\rm{l}}^{-2}+L_{\rm{s}}^{-2}\right)^{-1/2}\,.
\end{equation}
The corresponding function gives a fit line in
Fig.~\ref{fig:recDistribution}(a), which by the eye cannot be
distinguished from the solid one shown. This finding will become
useful later when discussing rotational effects in
Sec.~\ref{sec:rotation} and extensions of the theoretical treatment in
Sec.~\ref{sec:conclusions}.

\section{Additional rotational diffusion}
\label{sec:rotation}

For non-circular shapes of molecules, rotational effects can affect
the RTD and hamper an accurate determination of translational diffusion
coefficients, as described in the previous sections. On the other hand,
one may utilize the modifications of the RTD to quantify rotational
dynamics. To get insight in these effects, we consider here a simple
model of discrete rotational moves that occur independent of
translational moves (no rotation-translation coupling).  In this model,
rectangular-shaped molecules with aspect ratio $\alpha$, as considered
in Sec.\,\ref{sec:molShape}, perform transitions between $n$ possible
orientations separated by an angle $\Delta_\varphi=2\pi/n$. The
transitions occur between neighboring orientations around the molecule
center with a constant rate $w_\varphi =
D_\varphi/2\Delta_\varphi^2$, where $D_\varphi$ is the rotational
diffusion coefficient. In the following we use $n=10$, i.e.\
$\Delta_\varphi=\pi/5$ and molecules of size $10\times5$ in all
simulations.

\begin{figure}[t]
\includegraphics[width=8cm]{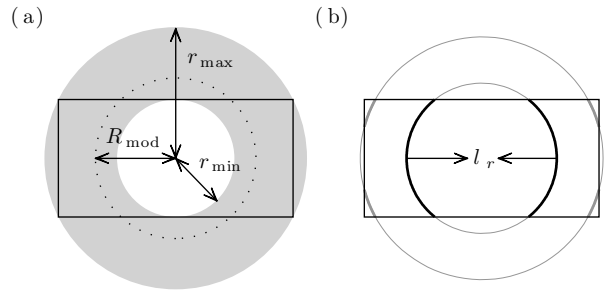}
\caption{(a) Classification of distances between the molecule center
  (center of figure) and probe tip: In the circular blank core area,
  rational moves cannot influence the signal, while in the shaded area
  rotational moves are relevant. The dotted line marks the circle with
  radius $R_{\rm mod}$ assigned to the rectangle [cf.\
  Eq.~(\ref{eq:rmod})]. (b) Origin of detection lines in the case of
  pure rotational diffusion: If the center of a rectangular molecule
  sets the origin of a co-rotating coordinate frame, rotational
  diffusion leads the (tip) probe to diffuse along concentric circles
  (thin lines).  The intersection set of these circles with the
  rectangle yields arcs that define detection lines of length $l_r$
  (bold lines).}
 \label{fig:sketch}
\end{figure}

We first study only rotational movements of a single molecule at a
fixed distance $r$ from the tip. For the analysis of this situation,
it is convenient to consider the equivalent problem of a fixed
molecule center and a tip performing jumps of size $r\Delta_\varphi$
on a concentric circle around the center. Clearly, if $r<r_{\rm
  min}\equiv L_{\rm s}/2$ [see inner circle in Fig.~\ref{fig:sketch}(a)],
the signal is always ``on'', while for $r>r_{\rm max}\equiv
(L_{\rm s}^2+L_{\rm l}^2)^{1/2}/2$ [see outer circle in Fig.~\ref{fig:sketch}(a)],
the signal is always ``off''.

In the regime $r_{\rm min}<r<r_{\rm max}$, a signal alternating
between ``on'' and ``off'' states can be obtained. To derive the RTD
$\Psi_{\rm rot}(t|r)$ for a given $r$ in this regime, we note that the
``on''-periods correspond to time intervals, where the tip is located
on certain arcs of the concentric circle with radius $r$. As sketched
in Fig.~\ref{fig:sketch}(b), two opposing arcs of equal length are
present if $r<L_{\rm l}/2$ (and $r>r_{\rm min}$), while for $r>L_{\rm l}/2$ (and
$r<r_{\rm max}$), four equivalent arcs close to the corners of the
rectangle appear. In analogy to the detection areas considered before
for the translational diffusion, the arcs form detection lines with
length $l_r$ given by
\begin{equation}
 l_r=\left\{\begin{array}{ll}
     2\,r\,\mathrm{asin}\left(\frac{L_{\rm{s}}}{2r}\right)\,, 
& r_{\rm min}<r\le\frac{L_{\rm l}}{2}\,,\\[1ex]
      r\left[\mathrm{asin}\left(\frac{L_{\rm{s}}}{2r}\right)
      -\mathrm{acos}\left(\frac{L_{\rm{l}}}{2r}\right)\right]\,, 
& \frac{L_{\rm l}}{2}<r<r_{\rm max}\,.
\end{array}\right.
\label{eq:lr}
\end{equation}

For jump lengths $r\Delta_\varphi$ much smaller than $l_r$, we can,
for calculating the RTD, consider the problem of a diffusing tip on a
line of length $l_r$ with absorbing boundaries. The initial
distribution is concentrated on two points at distance
$r\Delta_\varphi$ from the boundaries. After determining the
respective one-dimensional diffusion propagator $p(x,t)$, the first
passage time distribution $\Psi_{\rm rot}(t|r)$ follows from
$\Psi_{\rm rot}(t|r) = -\partial_t\int_0^{l_r} {\rm d}x\,
p(x,t)$, yielding
\begin{equation}
  \Psi_{\rm rot}(t|r)=
\frac{4D_\varphi}{\pi}
\displaystyle\sum_{n=0}^\infty 
\frac{q_n^2\sin\bigl(q_n\Delta_\varphi)}{(2n+1)}
\exp(-q_n^2D_\varphi t)\,,
\end{equation}
where $q_n=q_n(r)=(2n+1)\pi r/l_r$. After averaging over the area
of all positions between
$r_{\rm{min}}$ and $r_{\rm{max}}$, taking into account
that 2 equivalent arcs exist for $r_{\rm{min}}<r\le L_{\rm l}/2$ and four
equivalent arcs for $L_{\rm l}/2<r<r_{\rm{max}}$ 
the RTD
\begin{align}
\Psi_{\rm{rot}}(t) &= \mathcal{N}^{-1}\Biggl[
\int_{r_{\rm{min}}}^{L_{\rm l}/2}\mathrm{d}r\,r\,
\Psi_{\rm{rot}}(\mathbf{r},t)\nonumber\\
&\hspace{6em}
{}+2\int_{L_{\rm l}/2}^{r_{\rm{max}}}\mathrm{d}r\,r\,\Psi_{\rm{rot}}(\mathbf{r},t)
\Biggr]
\label{eq:psiROT}
\end{align}
is obtained, where $\mathcal{N}=\pi(L_{\rm s}^2+L_{\rm l}^2)/4$ is the
normalization factor. To account for the effect of the finite jump
length for detection lines with small length $l_r$, we had to deal
with a rather complex situation with, among others, very small numbers
of just 1-2 tip positions, whose precise location in turn depends on
$r$ and $\Delta_\varphi$. After averaging over $r$ these effects of
the discreteness of the jump length are, however, washed out.

In the presence of both rotational and translational diffusion, it is
difficult to obtain exact analytical results for the RTD, because the
problem cannot be described as a diffusion problem with a
time-independent geometry of the absorbing boundary. Fortunately, in
the situation, where rotational diffusion is relevant in the RTD, the
results obtained for pure translation and pure rotation are sufficient
to account for the overall behavior, as explained in the following.

The signal can turn from ``on'' to ``off'' due to rotational moves
only if the molecule center is in the shaded area in
Fig.~\ref{fig:sketch}(a).  A typical arc in this area has an angle of
about $\pi/4$ to $\pi/2$ [cf.\ Fig.~\ref{fig:sketch}(b)], which
results in a typical time $\sim
\tau_\varphi\equiv(\pi/4)^2/2D_\varphi$ for the molecule to leave the
detection area by rotation. The typical time for a molecule center to
traverse the shaded area in Fig.~\ref{fig:sketch}(a) is $\sim\tau_{\rm
  tr}\equiv(L_{\rm l}/2-L_{\rm s}/2)^2/4D=(\alpha-1)^2L_{\rm s}^2/16D$. Hence, if
$\tau_\varphi\gg\tau_{\rm tr}$, the decay of the RTD should be
governed by translational diffusion as described in
Sec.~\ref{sec:molShape}. On the other hand, if
$\tau_\varphi\ll\tau_{\rm tr}$, the rotational diffusion should become
significant. It governs the RTD for short times $t\ll\tau_{\rm tr}$,
while for $t\gtrsim\tau_{\rm tr}$ the dominant events are those, where
the molecule center enters the ``core area'' $r\le r_{\rm min}$ and
leaves it by translational diffusion. Accordingly, the RTD becomes
decomposable into one part given by pure rotational diffusion, i.e.\
Eq.~(\ref{eq:psiROT}), and a second part given by pure translational
diffusion, i.e.\ Eq.~(\ref{eq:psiRTD}) with $R=r_{\rm min}$.

\begin{figure}[t]
 \centering
 \includegraphics[width=8cm]{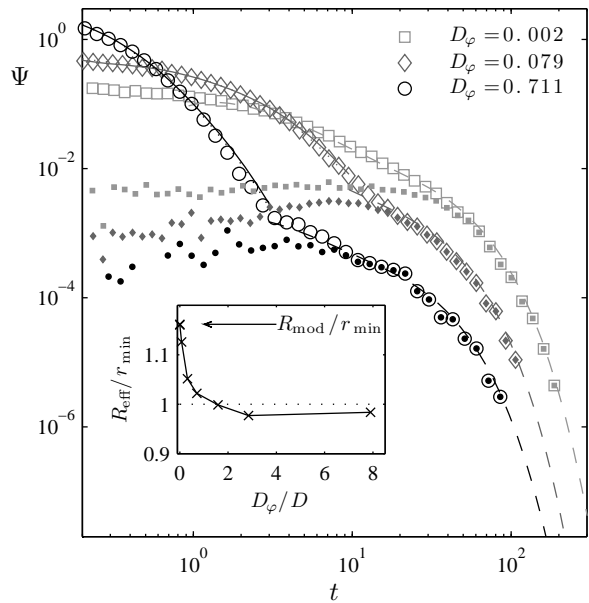}
 \caption{(a) Simulated RTDs for three different $D_\varphi$ (open
   symbols). Dashed lines correspond to Eq.~(\ref{eq:psiRTD}) for
   translational diffusion with $R=R_{\rm eff}$ (see main text). Solid
   lines correspond to Eq.~(\ref{eq:psiROT}) for rotational
   diffusion. The small solid symbols refer to distributions of
   residence times, which belong to trajectories, where the molecule
   center has at least once entered the core area shown in
   Fig.~\ref{fig:sketch}(a). The inset shows the dependence of the
   effective radius $R_{\rm eff}$ on $D_\varphi/D$.}
 \label{fig:rotation}
\end{figure}

Results of KMC simulations shown in Fig.~\ref{fig:rotation} confirm
these considerations. In this figure, representative RTDs in the
presence of both rotational and translational diffusion are displayed
for a rectangular molecule of size $10\times5$ for various
$D_\varphi/D$ ($=D_\varphi/Da^2$) ratios. For
$D_\varphi/D=8\times10^{-3}$ ($\tau_\varphi/\tau_{\rm
  tr}=25\gg1$) rotational diffusion indeed has no influence, and
the KMC data can be well described by Eq.~(\ref{eq:psiRTDrec}) (dashed
line).  For $D_\varphi/D=2.8$ ($\tau_\varphi/\tau_{\rm tr}=0.07\ll1$),
a double shoulder characterizes the distribution. This reflects the
separation into the two time regimes governed by rotational and
translational diffusion, as demonstrated by the curves corresponding
to Eq.~(\ref{eq:psiROT}) (solid line) and to Eq.~(\ref{eq:psiRTD})
with $R=r_{\rm min}$ (dashed line). The relative weight of the two
contributions was determined in the following way: By making the
ansatz $\Psi(t)=B_1\Psi_{\rm rot}(t)+B_2\Psi_{\rm circ}(t)$, the
coefficient $B_2$ was first determined by fitting $\Psi(t)\sim
B_2\Psi_{\rm circ}(t)$ to the KMC data in the long-time regime [with
$\Psi_{\rm circ}(t)$ taken from Eq.~(\ref{eq:psiRTDII})]. Thereafter,
the coefficient $B_1$ followed from the normalization of $\Psi(t)$.  The
small full symbols in Fig.~\ref{fig:rotation} represent the
distributions of residence times, during which the molecule center has
entered the core area. These distributions are normalized to the overall
fraction of the corresponding events. Their good agreement
with the long-time behavior of the simulated data is a further proof
that this regime is dominated by translational diffusion in the core
area.

Generally, the influence of the rotational motion on the long-time
regime can always be captured by defining an effective radius $R_{\rm
  eff}$, which follows from fitting the exponential decay in the
long-time regime to Eq.~(\ref{eq:psiRTDII}). The behavior of this
effective radius as a function of $D_\varphi/D$ is shown in the inset
of Fig.~\ref{fig:rotation}. For $D_\varphi/D=0$ (no rotational
diffusion), $R_{\rm eff}=R_{\rm mod}$, as discussed at the end of
Sec.~\ref{sec:molShape}.  With increasing $D_\varphi/D$, $R_{\rm eff}$
decreases and rapidly approaches $R_{\rm eff}=r_{\rm min}$.  When
$R_{\rm eff}=r_{\rm min}$, the rotational moves are so fast that, if a
molecule center leaves the core area, the signal will almost
immediately be turned off. The radii $R_{\rm eff}>r_{\rm min}$ for
$D_\varphi/D\lesssim1$ can be assigned to an effective circular
detection area, which takes into account that a molecule center, when
leaving the core area, typically diffuses over a certain effective
distance before the signal is turned ``off'' because of a rotational
move.

Let us finally address how our findings can be applied to extract
rotational and translational diffusion coefficients if both types of
molecular motion are present. As discussed above, if
$\tau_\varphi\gg\tau_{\rm tr}$ [$D_\varphi/D\ll (L_{\rm l}/2-L_{\rm
  s}/2)^{-2}$] only translational diffusion coefficients can be
determined from the RTD. When a double shoulder shows up in the RTD
for $\tau_\varphi\lesssim\tau_{\rm tr}$ [$D_\varphi/D\gtrsim (L_{\rm
  l}/2-L_{\rm s}/2)^{-2}$], the rotational diffusion coefficient
should also be determinable.  In fact, using a Levenberg-Marquardt
nonlinear fitting of Eq.~(\ref{eq:psiROT}) to the shoulder in the
short-time regime, we obtain $\tilde D_\varphi=0.081$ for the diamonds
and $\tilde D_\varphi=0.69$ for the circles, which agree well with the
respective input values $D_\varphi=0.079$ and $D_\varphi=0.71$.
Simultaneously, by fitting Eq.~(\ref{eq:psiRTDII}) to the shoulder in
the long-time regime, $\tilde D$ values are determined. For this
analysis, one can choose $R_{\rm eff}=r_{\rm min}$ first. If the
resulting $\tilde D_\varphi/\tilde D$ turns out to be larger than one,
$\tilde D$ should have a reliable value. However, if $\tilde
D_\varphi/\tilde D\lesssim1$, the $\tilde D$ value is underestimated
because the effective radius $R_{\rm eff}$ is larger than $r_{\rm
  min}$. For our data in Fig.~\ref{fig:rotation}, we obtain $\tilde
D=0.23$ (diamonds) and $\tilde D=0.25$ (circles), which in connection
with the $\tilde D_\varphi$ values give $\tilde D_\varphi/\tilde
D=0.36$ and $\tilde D_\varphi/\tilde D=2.76$, respectively. Indeed,
$\tilde D$ for the circles agrees well with $D=1/4$, while $\tilde D$
for the diamonds is by 8\% smaller.\cite{comm:reff} In practice, it
will generally be unlikely to encounter this deviation, because of the
narrow regime $D_\varphi/D\lesssim1$, where $R_{\rm eff}$ is larger
then $r_{\rm min}$.  If the problem is nevertheless present, the
experimentalist may shift the $D_\varphi/D$ value to the favorable
regime by changing the temperature.

\section{Conclusions and discussion}
\label{sec:conclusions}

Analysis of signal fluctuations from a locally fixed probe is a
powerful means to determine diffusion coefficients.  In this work we
concentrated on the determination of diffusion coefficients of
molecules on surfaces, but the concepts can also be useful in other
fields such as, e.g., single-molecule fluorescence
microscopy.\cite{Petrov/Schwille:2008,Zumofen/etal:2004}

The first focus of our treatment was on the implications of molecular
shapes on the RTD. Since typical molecules used in experiments, as
shown in Fig.~\ref{fig:footprints}, can have rectangle-like geometries
$L_{\rm l}\times L_{\rm s}$ with aspect ratios $\alpha=L_{\rm l}/L_{\rm s}$ significantly
larger than one, it was necessary to extend our former treatment, that
was limited to molecules with a circular
shape.\cite{Hahne/etal:2013} The exact solution
Eq.~(\ref{eq:psiRTDrec}) for such rectangular shapes was
successfully applied to data obtained from KMC simulations, which
served as surrogate for experimental results. It was shown that for
$\alpha\gtrsim1.5$, an application of the RTD for circular
shapes, with $R$ taken as the gyration radius, yields erroneous
results. However, when taking the modified radius defined in
Eq.~(\ref{eq:rmod}), the solution for circular shapes can also be
used. It gives a very good approximation to the exact
solution.

As discussed in the introduction, the RTD is the most favorable
variant for analyzing the signal fluctuations, and we therefore
focused our treatment to it here. Nevertheless, in some situations it
may be helpful to analyze also the interpeak time distribution (ITD)
or autocorrelation function (ACF). For the ACF, the consideration of
shapes different from circular ones can be readily accounted for by
defining corresponding detection functions, see
Ref.~\onlinecite{Hahne/etal:2013}.

For the interpeak times, on the other hand, we were not able to derive
their distribution. One would need the diffusion propagator for the
outer domain of an absorbing rectangle.  A solution for this outer
problem is difficult and we did not succeed to derive a closed form or
to find a derivation in the literature. However, with the finding that
the RTD for rectangular shapes can be well approximated by the RTD for
circular shapes when introducing an appropriate radius, one can follow
a corresponding route to find an approximate solution for the ITD. At
long-times, the ITD is governed by exchange processes of different
molecules, yielding an exponential decay with characteristic time
$\propto d^2/D$, where $d$ is the mean intermolecular distance.  For
rectangular-shaped molecules, $d=(L_{\rm l}L_{\rm s}/\theta_0)^{1/2}$, where
$\theta_0$ is the coverage of the freely diffusing molecules. In
practice, this coverage can be extracted from the signal as the ratio
of the ``on-periods'' to the total observation time $T$, that is $\theta_0
= \sum_i t_i/T$.  For circular shapes, $d=(\pi R^2/\theta_0)^{1/2}$.
Comparing the corresponding characteristic times $d^2/D$, the
effective radius
\begin{equation}\label{eq:rcov}
R_{\rm{cov}}=\left(\frac{L_{\rm l}L_{\rm s}}{\pi}\right)^{1/2}
\end{equation}
is obtained. Inserting $R_{\rm{cov}}$ for $R$ in the exact solution
for circular-shaped molecules [Eq.~(10) in
Ref.~\onlinecite{Hahne/etal:2013}], one can check whether this
provides a good approximation of the ITD of rectangular-shaped
molecules.  By comparison with data obtained from our KMC simulation,
we indeed found good agreement.

The second focus of our treatment was on the implications of
rotational diffusion on the RTD by considering as a first step a model
of uncoupled rotation and translation. These implications become the
more important the larger the aspect ratio is. For pure rotational
diffusion, the analytical expression Eq.~(\ref{eq:psiROT}) was derived
by considering diffusion along circular arcs, and by averaging over
all possible tip-molecule distances. Combining this solution with the
solution for pure translational diffusion, we succeeded to describe
the behavior of the RTD when both rotational and translational
diffusion are present. We showed, that when the ratio $D_\varphi/D$ of
the rotational to the translational diffusion coefficient is larger
than one, two shoulders appear in the RTD. From those, $D_\varphi$ and
$D$ can be extracted separately.  For $D_\varphi/D\lesssim1$, an
effective radius, varying between $r_{\rm min}=L_{\rm s}/2$ and $R_{\rm
  mod}$, needs to be taken into account. By employing KMC simulations
we demonstrated how the theoretical approach can be applied.

Also for rotational diffusion we can discuss the implications on the
ACF and ITD. In the case of the ACF, one can straightforwardly
generalize (i) the detection functions for the angle degrees of
freedom and (ii) the free diffusion propagator, which becomes simply
the product of the free propagators for rotational and translational
diffusion in the uncoupled case. The derivation of the behavior of the
ITD can be done in close analogy to the RTD treatment. When deriving
the part arising from pure rotation, one needs to replace the inner
arcs [thick lines in Fig.~\ref{fig:sketch}(b)] with the outer arcs
[thin lines in Fig.~\ref{fig:sketch}(b)]. Combining the corresponding
solution with the approximate solution for pure translation (see
discussion above), an effective radius now appears that varies between
$R_{\rm{area}}$ (for $D_\varphi/D\to0$) and $r_{\rm
  max}=[(L_{\rm l}/2)^2+(L_{\rm s}/2)^2]^{1/2}$ (for $D_\varphi/D\gtrsim1$).
Using the KMC simulations we again checked that this procedure gives
good estimates for both diffusion coefficients $D_\varphi$ and $D$.

The model of uncoupled rotation and translation with a fixed jump
angle $\Delta_\varphi$ is certainly a strong simplification and
further work is necessary to account for non-uniform jump angles,
rotation centers displaced from the molecule center,
rotation-translation couplings, etc. Irrespective of these
complications, the possibility to simultaneously determine rotational
and translational diffusion coefficients is a promising perspective
that calls for an experimental verification.

\begin{acknowledgments}
  The authors thank M.~Sokolowski for helpful discussions and
  A.~K\"uhnle for a critical reading of the manuscript.
\end{acknowledgments}


\begin{thebibliography}{22}
\bibitem [{com()}]{comm:reff}%
  \BibitemOpen
  \href@noop {} {}\bibinfo {note} {{This deviation is relatively small, because
  $R_{\rm eff}$ is close to $r_{\rm min}$ for this example. Generally we found,
  that when the double shoulder can be clearly identified, $R_{\rm eff}$ and
  $r_{\rm min}$ do not differ much.}}\BibitemShut {Stop}%
\end{thebibliography}

%

\end{document}